\documentclass[runningheads,a4paper]{llncs}

\usepackage{amssymb}
\usepackage{graphicx}
\usepackage{subfigure}
\usepackage{url}
\urldef{\mailsa}\path|{alfred.hofmann, ursula.barth, ingrid.haas, frank.holzwarth,|
\urldef{\mailsb}\path|anna.kramer, leonie.kunz, christine.reiss, nicole.sator,|
\urldef{\mailsc}\path|erika.siebert-cole, peter.strasser, lncs}@springer.com|
\newcommand{\keywords}[1]{\par\addvspace\baselineskip
\noindent\keywordname\enspace\ignorespaces#1}

\usepackage{float}
\usepackage{graphicx}

\usepackage{indentfirst}
\usepackage{amsmath}
\usepackage{slashbox}
\usepackage{clrscode}
\newtheorem{myDef}{Definition}

\begin{document}

\mainmatter  

\title{Stable Matching with Incomplete Information in Structured Networks}

\titlerunning{Stable Matching in Structured Networks}

%
%
\author{Ying Ling$^1$, Tao Wan$^2$ and Zengchang Qin$^1$$^*$
}
\authorrunning{Ling, Wan and Qin}

\institute{$^1$Intelligent Computing and Machine Learning Lab\\
      School of ASEE, Beihang University, Beijing, 100191, China\\
      $^2$School of Biological Science and Medical Engineering\\
      Beihang University, Beijing, 100191, China\\
     \tt  $^*$zcqin@buaa.edu.cn \\
}

%
%

\toctitle{Stable Matching in Structured Networks}
\tocauthor{Ling, Wan and Qin}
\maketitle

\begin{abstract}
In this paper, we investigate stable matching in structured networks. Consider case of matching in social networks where candidates are not fully connected. A candidate on one side of the market gets acquaintance with which one on the heterogeneous side depends on the structured network. We explore four well-used structures of networks and define the social circle by the distance between each candidate. When matching within social circle, we have equilibrium distinguishes from each other since each social network's topology differs. Equilibrium changes with the change on topology of each network and it always converges to the same stable outcome as complete information algorithm if there is no block to reach anyone in agent's social circle.
\keywords{Stable Matching; Structured Networks; Social Circle; Equilibrium}
\end{abstract}

\section{Introduction}

Stable matching can be best explained
by the example of marriage and thus also known as the \emph{stable marriage problem} (SMP).
That aims to find a stable matching between two equally sized sets of elements given an ordering of preferences for each element.
A matching is a bijection mapping from the elements of one set to the elements of the other set.
Two sets can be illustrated as an equal number $n$ of men and women, in which every man ranks the $n$ women according to how desirable each is to him, without ties. Similarly, every woman ranks the $n$ men based on their willingness \cite{Gale1985}. Ideally, a perfect match would pair every man with the woman he likes best and vice versa. Clearly the preferences expressed by men and women rarely allow for a perfect match. But we can go for a \emph{stable match}.
A stable match is a match such that there is no man and woman that both like each other better than their current partners. When a match is stable, all couples are static: a man tempted to abandon his current partner for another woman he ranks higher will be rebuffed, since that woman ranks her partner higher than him. We can start with random matching, exchange the unstable pairs by switching their partners until no pairs have motivation to change, so the matching is stable. Such a solution os called Gale-Shapley algorithm \cite{GS1962}.

The classical Gale-Shapley algorithm assumes that all the information is public known, each agent is with complete
information. Few works have been reported to study the stable matching with incomplete information[cite].
In the real-world matching problems, information is always limited. Just like the fact that it is infeasible for every man knows every woman and vice versa in the real world. Such incompleteness may impact the matching results significantly.
In this paper, we assume the acquaintance between agents is incomplete and it can be modeled by a network. A fully connected
network indicates ideal complete information.
We are living in communities through social connection.
In standard stable marriage problem, all the people are matched but in our research, we assume some agents could stay unmatched.

The remainder of the paper is structured as follows.
We propose a new matching algorithm inspired by graph theory in Section 2. Section 3 gives theoretical analysis of graphical model and its influences on matching outcome. Section 4 characterizes the networked matching process in details. In Section 5 we introduces the equilibrium results and the intrinsic reason behind the results of four social networks. Finally, the conclusions and future work are given in the last section.

\section{Network Structure}

A social network is a social structure made of a set of agents and a set of the dyadic ties between them. An agent could be an individual or an organization.
The network structure affects agents' social behaviors, so as to determine social phenomena in macro-level. In this paper, we mainly consider the following well-studied networks: Scale-free networks (BA model) \cite{BA}, random networks (ER model) \cite{ER}, small world networks (WS model) \cite{strogatz2001} and nearest-neighbor Coupled Network (NCN model). The reason for choosing these four structures is because they are well-used representative social networks \cite{Li2014}.
The analysis of the topology of complex networks is the key to study the dynamic progress of propagation dynamics, network synchronization, traffic flow and node game \cite{newman2003}. This research provides new approaches to investigate problems such as group consensus making, networked bargaining and trading strategies.

In graph theory, a network can be viewed as a graph $G=(V,E)$, which is composed of a set of nodes $V$ and edges $E$. Node number $N=|V|$, where $|.|$ represents the cardinality, and the number of edges is $M=|E|$. Each side of the $E$ has two nodes in $V$ and corresponding to it.
If not mentioned particularly, undirected network and unauthorized network are \emph{not} within our consideration.
We only consider the simple network - there is no self-loops or repeated edges.

\subsection{Four well-used Networks}

BA model have been proposed as a model that reflects how social networks are formed, particularly online \cite{Kitsak2007}. The network is seeded with two random links. After this, to add a link, we choose a node randomly and consider the links it could add to the graph. Each link is given a weight equal to the degree of the target node it connects to, and a link is chosen in proportion to these weights.

In the $ER (n,p)$ model, a graph is constructed by connecting nodes randomly with probability $p$ independent from every other edge. This is a baseline process where we add a link chosen uniformly from those links that do not already exist in the graph \cite{Gomez-Gardenes2006}.

As a transition from the completely regular network to the completely random network, the introduction of a little randomness into regular network can generate a network with small world characteristics, now known as WS small-world network model \cite{Latora2001}.

Nearest neighbor-coupled network of periodic boundary conditions formed a ring of $N$ vertex, where each node and its neighbors around are connected, $K$ is an even number. The most important feature of this kind of network is that the topological structure is determined by the relative position between the nodes, the network topology may also occur switch when the position of the nodes changes.

\subsection{Network Topology}

The topology of the network decides the dynamics of the network, two parameters characterizing complex network topology
were well used in \cite{wang2011}: degree distribution, the average path length (APL).
The degree $k_i$ of the node $i$ refers to the number of edges connected to the node $i$. The average degree of all the modes in the network is denoted as $\overline{k}$:
\begin{equation}
\overline{k}=\sum_{i=1}^N k_i /N
\end{equation}
The probability distribution function of the degree is generally denoted as $p(k)$, regarding to the probability of a node connecting with $K$ nodes.

The distance between two nodes $i$ and $j$ $(l(i,j))$ is defined as the number of edges in the shortest path connecting the two nodes. Here we use Dijkstra's algorithm \cite{dijkstra1959} to calculate the shortest distance between two nodes. The average path length of the network is defined as the average value of the distance between any two nodes.
\begin{equation}
\overline{l}=\frac{\sum_{i>j}l(i,j)}{{N(N-1)}/{2}}
\end{equation}
The average path length $\overline{l}$ of four models are listed in Table \ref{tab:APL} \cite{wang2011}.

\begin{table}[!h]
\caption{Average path length of four well-used network models.}
\begin{center}
\begin{tabular}{|c|c|c|c|c|}
\hline
Network Model & \quad NCN \quad  & \quad  ER \quad & \quad WS \quad &  \quad   BA \quad\\
\hline
APL & \quad  $\overline{l} \varpropto N$ \quad &  \quad   $\overline{l} \thickapprox \frac{ln N}{ln \overline{k}}$ \quad
& \quad  $\overline{l}=\frac{\sum_{i>j}l(i,j)}{{N(N-1)}/{2}}$ \quad  & $\quad  \overline{l} \varpropto \frac{\log N}{\log \log N}$ \quad \\
\hline
\end{tabular}
\label{tab:APL}
\end{center}
\end{table}

According to the degree distribution, complex network can be roughly divided into two big categories of scale-free networks and homogeneous networks[cite]. Homogeneous networks include nearest-neighbor coupled network, random network and small world network \cite{watts2003}. BA is scale-free network.
In this section we describe a few models we use to create social network graphs. Network formation is determined by various growth processes that describe how a link is added to an existing graph.

\section{Matching Algorithms for Networks}

In social and economic interactions, including public goods provision, job search, political alliances, trade and partnership, an agent's vision and experience depends on his or her neighboring relations. Neighboring relations can form a
network whose structure decides the direct interaction. In the real-world, regional limitation and attenuation of information flow help to make social circle for human beings. This truth inspired us to learn structured network in order to understand how the changes in network structure will reshape the matching outcomes.

\subsection{Social Circle}

\begin{myDef}
(Social Circle).  An individual's Social Circle reveals who are within the contact range is defined by a maximum depth $(dep)$ he or she could on reach through another one. From the point of view of a graph, we means $d(i,j)$ less than a certain given depth for mutual acquaintance.
\begin{equation}
sc(i)=
\begin{cases}
1& \text{$l(i,j)\leq dep$}\\
0& \text{$l(i,j)>dep$}
\end{cases}
\end{equation}
\end{myDef}

From the definition, it's obvious that when $l(i,j)> dep$ means vertex $j$ is \emph{not} in vertex $i$'s social circle and and vice versa.

Compared to standard stable matching problem, the structure of each agent's social circle may have a great influence on
its final matching results. For example, if all potential partners in one's social circle were chose by other competitive players, he(she) would be left unmatched and turns out no utility at all. This cannot happen in G-S matching.
When considering structured networks when using the classic G-S matching algorithm, the connectivity of each network influences its efficiency in connected agents.

\begin{myDef}
(Network Connectivity).  Network Connectivity of a network $\phi$ refers to the proportion of the number of paths whose length less than the maximum depth $(dep)$ to the number of all possible paths in a network.
\begin{equation}
\phi=\frac{count(l\leq dep)}{{N(N-1)}/{2}}
\end{equation}
\end{myDef}

It directly determines the number of people in someone's social circle. The more participation, the higher utility we can have, while connectivity for the classical G-S algorithm always be considered $\phi_{GS}=1 $.

Actually, APL in each social network forms the difference in connectivity at the start.
The distribution of the shortest path length of the random network (ER) obeys Poisson distribution:
\begin{equation}
P(X=d)=\frac{\lambda^d e^{-\lambda} }{l!} (l=1,2,3...)\end{equation}
Where $\lambda$ is the average path length.
Then, the connectivity can be formulated by $dep$, $d$ and $\overline{l}$.

\begin{equation}
\phi=\int^{dep}_0\frac{\lambda^l e^{-\lambda}}{l!}{\rm d}(l)
\end{equation}
Through theoretical derivation, the average path length of random network (ER) is negatively correlated with connectivity (i.e., $\lambda \uparrow  \rightarrow  \phi \downarrow$).

From Section 2, we have seen that two connected nodes in the NCN model are adjacent ones. Two nodes are connected in accordance with a certain probability in ER model. WS model adds some randomness into regular network, while BA model emphasizes node weight.

One node is to reach different kind and different number of other nodes in four social networks, leading to our research on stable matching problem based on structured networks. As to degree distribution, it's easy to understand that the more people participate in the match, the more choices people have, the more the utility will comes out.

As theoretical analysis of networks' topology shown, APL is a basic element determining average utility in agreement with external characteristics: differences in ways of networks connection and the degree of distribution.
Relationship of APL and connectivity in the rest three models will not repeat them here. In the following part they will have empirical support.

\section{Matching in Structured Graphs}

There is a large literature of studying the matching models for market analysis with two-sided heterogeneity, such as the matching problems of students and schools, husbands to wives, and workers to firms.\footnote{See Roth and Sotomayer (1990) for a survey of two-sided matching theory} Typical analysis in literatures assumes that the agents have complete information, and then examines stable outcomes.
The assumption of complete information makes the analysis tractable but stringent. In this paper, agents on one side of the market cannot know all the candidates only knows the ones within his(her) social circle.

Let us reconsider the problem in the marriage setting as well:
There is a finite set of women, $I$, with an individual woman is denoted by $i \in I$. There is also a finite set of men, $J$, with an individual man is denoted by $j \in J$.

A matching pair function $p$ : $I\rightarrow J $, one-to-one on $p\ (i)$, that means woman $i$ matches with the man $p\ (i)$, where $p\ (i)= \varnothing$ means that woman $i$ is unmatched and $p^{-1}(j)= \varnothing $ means that man $j$ is unmatched.

Given a match between woman $i$ and man $j$, $S$ represents the ranking. woman $i$'s preference to man $j$ is recorded as $S_{i,j}^w$, while man $j$'s preference over woman $i$ is $S_{i,j}^m$. There is no necessary that $S_{i,j}^w=S_{i,j}^m$. In the G-S model, everyone could tell his or her absolute preference because of complete information. But in our model, we need emphasize that a woman $i$ must have the same preference at the same man $j$ in four different social networks if they know each other. For example, when woman $i$ make score for same man $j$, we should point that $S_{i,j}^{w,NCN}=S_{i,j}^{w,ER}=S_{i,j}^{w,WS}=S_{i,j}^{w,BA}$.

\begin{myDef}
(Utility of Agent).  Utility is a measure of the agent's needs and desires through decision-making process. An allocation $(i,j)$ consists of a matching pair function $p$ and ranking scheme $S$. Utility of each agent is associated with $S$.
\end{myDef}
Woman's utility is
\begin{equation}
u_i^w=S_{i,p(i)}^w\quad for\ i\in I
\end{equation}
\begin{equation}
s.t. \  l(i,p(i))\leq dep
\end{equation}
While man's utility is
\begin{equation}
u_j^m=S_{p^{-1}(j),j}^m\quad for\ i\in I
\end{equation}
\begin{equation}
s.t. \  l(p^{-1}(j),j)\leq dep
\end{equation}

To avoid trivial cases, we associate zero utility with unmatched agents, setting $u_{i \varnothing}=u_{\varnothing j}=0$. So the utility of a matching pair and the average utility of matching outcome are
\begin{equation}
U_{i,j}=\frac{u_i^w+u_j^m}{2}
\end{equation}
\begin{equation}
U=\frac{\sum_{ij} U_{i,j}}{{N}/{2}}
\end{equation}

We consider one-to-one matching (i.e. no polygamy), with incomplete preference lists. Men and women play different roles. The algorithm takes as input the lists of preferences of men and women. Throughout the algorithm, men and women are divided into two groups: those that are engaged, and those that are free (i.e. not yet or no longer engaged). Initially, all men and all women are free.

As long as the group of free men is non-empty, the algorithm selects at random one man $j$ from the group of free men. Man $j$ proposes to the woman whom he ranks highest among recognized women $i$ to whom he has never proposed before. One of three following scenarios may happen:
$-$ $i$ is free. In this case, $j$ and $i$ are engaged to each other and both move to the engaged group.

$-$ $i$ is already engaged to $j'$ but ranks $j'$ less than $j$. In this case, $i$ breaks her engagement to $j'$ and instead gets engaged to $j$. $j$ and $i$ join the engaged group, whereas $j'$ goes back to the group of free men.

$-$ $i$ is already engaged to $j'$ and ranks $j'$ ahead of $j$. In this case, $i$ stays engaged to $j'$ and $j$ stays in the group of free men.

\begin{myDef}
(A Stable Matching Outcome).  A stable matching outcome given a social circle means there is no woman-man combination $(i,j)$ such that \cite{Hoppe2011}
\begin{equation}
 u_{w(i),m(j)}>u_{w(i),m(p(i))}
\end{equation}
\begin{equation}
u_{m(j),w(i)}>u_{m(j),w(p^{-1}(j))}
\end{equation}
for all $(i,j)$ satisfying

\begin{equation}
d(i,j)\leq dep
\end{equation}
\end{myDef}

A proposed outcome that matches each man to a recognized woman is stable if there is no unmatched man-woman pair that could increase both their utility by matching with each other within their social circles.
\begin{table}[!h]
\begin{tabular}{l}
\hline
Find equilibrium algorithm for matching in BA network.\\
\hline
\hline
Input: a game structure($N$ and preference list) \\
Output: equilibrium\\
We have original completely connected network with few vertexes\\
add new vertex $x$ into $y$ old vertexes with possibility of $p_{x}=\frac{k_{x}}{\sum_{y}k_{y}}$\\
calculate shortest path between each agent using Dijkstra algorithm\\
form $N$ agents' social circle\\
while(man $j$ is free)\\
$\quad$    $i$ $\leftarrow$ $j$'s top woman in his preference list he never proposed to before\\
$\quad$      if $i$ is free\\
$\qquad $      (i,j) become a match\\
$\quad$      else $i$ have dated with $j'$\\
$\qquad $     if $i$ prefers $j'$ to $j$\\
$\qquad\quad$        $j$ stays free and propose to the next ranking woman $i'$ \\
$\qquad\qquad$        if $i'$ is beyond $j$'s social circle\\
$\qquad\qquad\quad$            $j$ will stay unmatched in this game\\
$\qquad\qquad\quad$             $j'$ start to repeat\\

$\qquad$      else $i$ prefers $j$ to $j'$\\
$\qquad\quad$        (i,j) become a match\\
$\qquad\quad$        $j'$ become free\\
\hline
\end{tabular}
\label{tab:AveSatis}

\end{table}

Comparing to complete information stability, in our model, the process stops when every free man has tried to propose to all the women he knows ending with some men and women might be unmatched. This algorithm ensures every structured network is experimented under the same conditions and we might observe four different matching equilibriums when $N$ and $k$ are fixed, resulting
\begin{figure}[!hbt]
\begin{centering}
\centerline{\includegraphics[height=8cm,width=13cm]{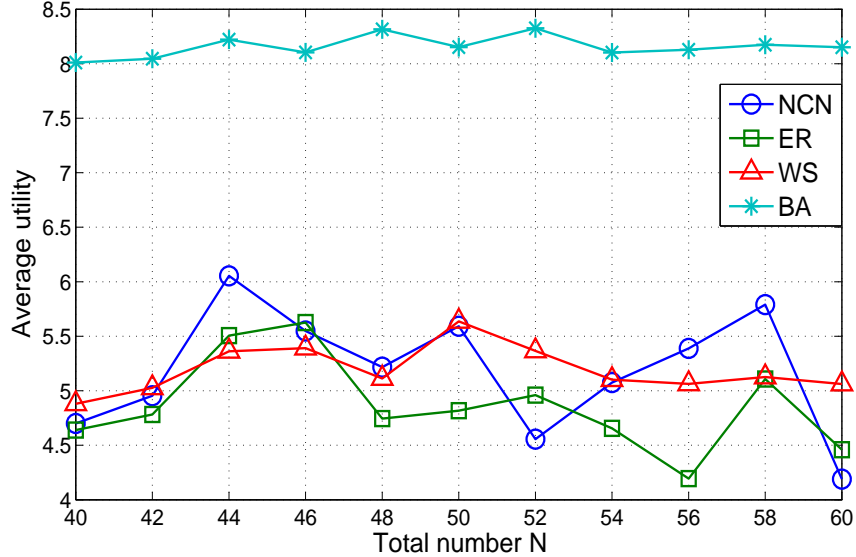}}
\caption{\  Relationship between total number and average utility. Four curves respectively describe the relationship between the matching population and average utility in each social model.}
\label{fig1}
\end{centering}
\end{figure}
from their differences in network topology.

\section{Experimental Studies}

As we have discussed in previous sections that network topology could explain the matching outcomes in structured networks. We investigated relatively small networks, with total number of nodes $N\leq100$. In each round, each player complete preference list over all potential partners and make score from one to ten associated with ranking:

$$S\in[1,10]$$

For a given matching game and preference list, we would run our various network formation algorithms and generate four kinds of graphs. In NCN model, each node is connected to its neighbors, this generate a sparse adjacency matrix which represents each agent's distance from another one. ER model is either of two closely related models for generating random graphs. In ER network, a graph is chosen uniformly at random from the collection of all graphs which have $n$ nodes and $M$ edges. This is a baseline process where we add a link chosen uniformly from those links that do not already exist in the graph. When NCN network is formed, WS can be generated by add some randomness. The original each edge of the network is randomly reconnected with the probability p, i.e., keep the endpoint of each edge unmoved, and the other endpoint is changed to a node randomly selected from the network. Actually, WS model a transition from NCN network to ER network. BA model introduce preferential attachment property into WS model. From a given connected network with $M$ nodes, each time a new node will be connected to the existing $m(m\leq M)$ nodes. But the probability when add this new node is related with the degree $k_{i}$ of the existing
\begin{figure}[!hbt]
\centering
\centerline{\includegraphics[height=8cm,width=13cm]{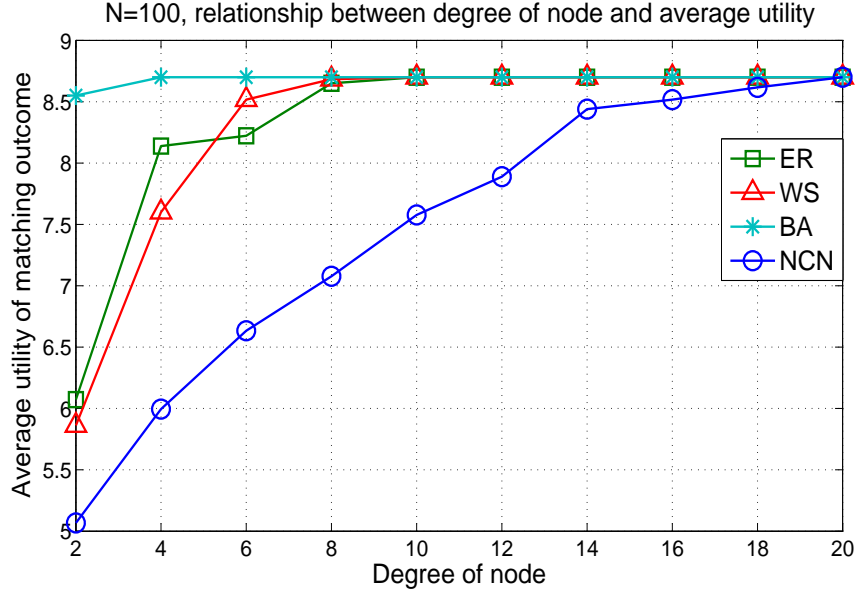}}
\caption{\   Relationship between degree of each node and average utility, four curves respectively express the corresponding average utility in each degree.}
\label{fig2}
\end{figure}
node $i$:
$$P_{i}=\frac{k_{i}}{\sum_{j}k_{j}}$$

While these networks are considerably smaller than the real networks, their scaling and topological features should be representative. We also have $dep=3$ to construct our model for the maximum depth between recognized participant. Four different models give out four different social circles for each agent.

In Fig. 1, it shows fluctuation near a certain level. That is, the total population plays no decisive role in matching results. But under different network, it implies that the utility of matching distinguishes form each other as well as the equilibrium.
We now show the average utility of four well-used network models under different population within degree of each node $k=2$ as Table \ref{tab:AveSatis}.

\begin{table}[!h]
\caption{The average utility of four well-used network models under different population within $k=2$.}
\begin{center}
\begin{tabular}{|c|c|c|c|c|}
\hline
\backslashbox{Population}{Network Model}  & \quad NCN \quad  & \quad  ER \quad & \quad WS \quad &  \quad   BA \quad\\
\hline
20 & \quad 4.600 \quad & \quad 4.620 \quad & \quad 5.610 \quad & \quad 8.000 \quad\\
40 & \quad 4.700 \quad & \quad 4.640 \quad & \quad 4.880 \quad & \quad 8.010 \quad\\
60 & \quad 4.189 \quad & \quad 4.459 \quad & \quad 5.062 \quad & \quad 8.151 \quad\\
80 & \quad 4.809 \quad & \quad 4.911 \quad & \quad 5.601 \quad & \quad 8.174 \quad\\
100 & \quad 4.742 \quad & \quad 4.958 \quad & \quad 4.970 \quad & \quad 8.265 \quad\\
\hline

\end{tabular}
\label{tab:AveSatis}
\end{center}
\end{table}

But, it cannot be ignored BA model has a higher level of average utility than other three models which developed within our prediction. This can be explained by the node connection mode of BA model itself. When we have opportunity to
contact with the relatively central player, participants are very likely to shorten the distance form another central one.

For the differences from model intrinsic properties. The number of participants in $N$ is fixed, the parameters of each social model changes to explore the reasons for the impact of the matching results.
\begin{figure}
\begin{minipage}[t]{0.5\linewidth}
\centering
\includegraphics[scale=0.33]{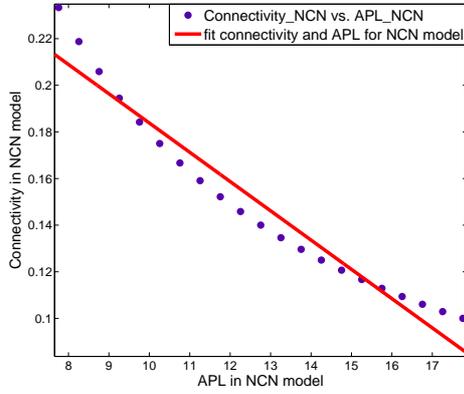}
\caption{connectivity vs APL for ER model \label{fig3}}
\end{minipage}
\hfill
\begin{minipage}[t]{0.5\linewidth}
\centering
\includegraphics[scale=0.33]{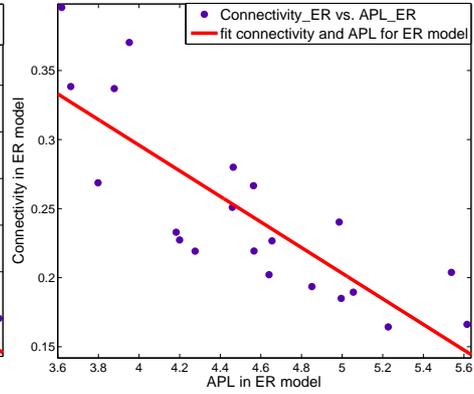}
\caption{connectivity vs APL for ER model\label{fig4}}
\end{minipage}
\end{figure}

\begin{figure}
\begin{minipage}[t]{0.5\linewidth}
\centering
\includegraphics[scale=0.33]{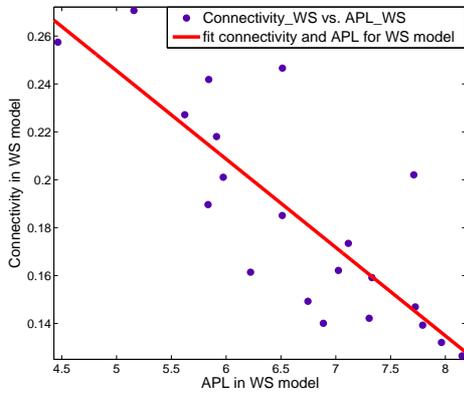}
\caption{connectivity vs APL for WS model \label{fig5}}
\end{minipage}
\hfill
\begin{minipage}[t]{0.5\linewidth}
\centering
\includegraphics[scale=0.33]{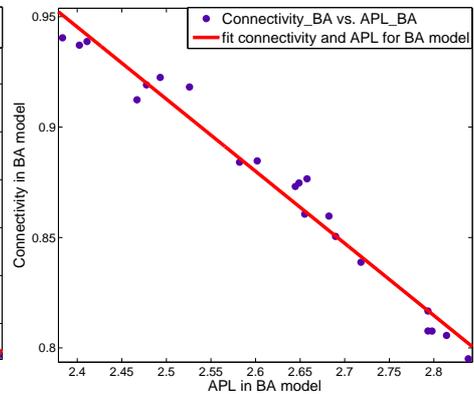}
\caption{connectivity vs APL for BA model\label{fig6}}
\end{minipage}
\end{figure}
Note that (see Fig. 2), the average utility of each model increases with the number of network degree. This result underlies matching outcome is related with network dynamic through topology.
Moreover, all four models' average utility converge to a certain value. That because everyone has made partnership of form $(i,j)$ which is complete information stable. At that point, every agent could get connected with anyone in the structured network even with defined social network.
In addition, with the increase of the degree, BA model gradually achieve the maximum utility, while NCN model is the worst one.

To understand why degree of each node determines a matching outcome and why four models differs from each other. We need to understand a higher connectivity of a model means each person have more choices because he or she know more people. So we get better matching results with higher average utility.

From the view of graph, four models distinguish from each other in intrinsic properties, degree of each node determines APL, which directly affects optional candidates number of each person. We get the empirical support shown in Fig. 3, Fig. 4, Fig. 5, Fig. 6. There exists negative correlation between APL and connectivity. The shorter the APL of a model, the higher the connectivity. The result underlines the importance of APL in understanding the dynamics of complex networks, and resolves the differences between four models.

Above all, in one network model, keep $N$ for a certain level, the higher degree of each node is, the shorter APL becomes. Then the higher connectivity is, the larger average utility is. Between four models, keep $N$ and degree remain the same, BA model have the shortest APL as well as the largest connectivity. So we always see BA model turns out best result.

There are many complex system in real world, such as the traffic network, the interpersonal relationship network, the WWW network, the neural network and so on. In recent research, many complex networks such as WWW networks and metabolic networks have Power Law distribution which now is called BA model \cite{Newman1999}. In simple terms, every day there are a lot of new web pages come to the WWW network. But in real network, a new node will have the greater probability to connect with the node possessing the higher degree. We call this ``Richer Get Richer" like Matthew effect. Besides of Internet, social contact also shows this trend in twitter, Facebook and so on. So far, BA model has its practical significance and the most efficiency in connection.

\section{Conclusion}

In this paper, we propose a stable matching algorithm by considering incomplete information in structured networks, where agents are not fully acquainted to each other. It emphasizes limited choice on marriage in reality, so we generate four well-used networks and define each agent's social circle to imitate real situation that there is no possibility to meet all people of opposite sex. We still use the same process to obtain all stable allocation when matching pairs, but end with every man has proposed to all woman within his social circle.

We find that different network structure makes each person's social circle different through its own average path length, leading to average utility of matching outcomes distinguishes from four models.
We start from a player(node) connects another node in four ways and forms four social networks. In these social networks, each player has a distance with rest players, we define $dep$ to divide who is on the choice list or not. Then we do matching algorithm only for recognized ones to get stable pairs, left some players be unmatched.
BA network plays as a better social organization to put a more productive matching outcome because of its efficient dynamics setting the connection between every one closer.

\section*{Acknowledgement}
This work is supported by the National Science Foundation of China Nos. 61305047 and 61401012.


\begin{thebibliography}{0}

\bibitem{Gale1985}Gale, D. and Sotomayor, M. (1985). ``Some remarks on the stable matching problem". Discrete Applied Mathematics, 11(3), 223-232.
\bibitem{GS1962}Gale, D. and Shapley, L.S. (1962). ``College Admissions and the Stability of Marriage". American Mathematical Monthly, 69, 9-14.

\bibitem{BA}Barabasi, A.L. and  Albert, R. (1999). ``Emergence of scaling in random networks. Science." Science, 286(5439), 509-512.

\bibitem{ER}Erdos, P. and Renyi, A. (1959). ``On Random Graphs. I". Publicationes Mathematicae, 6, 290-297.

\bibitem{strogatz2001}Strogatz, S.H. (2001). ``Exploring complex networks". Nature, 410(6825), 268-276.

\bibitem{Li2014}Li, Z. and Qin, Z. (2014). ``Impact of Social Network Structure on Social Welfare and Inequality".  Social Networks: A Framework of Computational Intelligence Springer International Publishing, 123-144.

\bibitem{newman2003}Newman, M. (2003). ``The structure and funktion of complex networks". Siam Review, 45(2), 167.

\bibitem{wang2011}Wang, X.C. and Jiang, Y.B. (2011). ``The influence of the randomness on average path length". Advanced Materials Research, 219-220, 791-794.
\bibitem{dijkstra1959}Dijkstra, E.W. (1959). ``A note on two problems in connexion with graphs". Numerische Mathematik, 1, 269-271.

\bibitem{watts2003}Watts, D. (2003). ``Six Degrees: the science of a connected age". W.w.norton and Co.inc.new York.

\bibitem{Gomez-Gardenes2006}Gomez-Gardenes, J. and Moreno, Y. (2006). ``From scale-free to erdos-renyi networks". Physical Review E Statistical Nonlinear and Soft Matter Physics, 73(5), 056124-056124.
\bibitem{Latora2001}Latora, V. and Marchiori, M. (2001). ``Efficient behavior of small-world networks". Physical Review Letters, 87(19), 198701.
\bibitem{Kitsak2007}Kitsak, M., Havlin, S., Paul, G., \emph{et al.}. (2007). ``Betweenness centrality of fractal and nonfractal scale-free model networks and tests on real networks". Phys.rev.e, 75(5), 96-96.
\bibitem{Hoppe2011}Hoppe, H.C., Moldovanu, B. and Ozdenoren, E. (2011). ``Coarse matching with incomplete information". Economic Theory, 47(1), 75-104.

\bibitem{Newman1999}Newman, M.E. (1999). ``Scaling and percolation in the small-world network model". Phys.Rev.E, 60(6), 7332-7342.
\end{thebibliography}
\end{document}